\documentstyle[12pt]{article}

\begin{document}
\begin{titlepage}

\flushright{To Appear: {\em Phys. Rev.} {\bf D15}, {\em Rapid Communications}}

\vspace{1in}

\begin{center}
\Large
{\bf Scale Factor Duality and Hidden Supersymmetry in Scalar--Tensor
Cosmology}

\vspace{1in}

\normalsize

\large{James E. Lidsey}

\normalsize
\vspace{.7in}

{\em Astronomy Unit, School of Mathematical
Sciences,  \\
Queen Mary \& Westfield, Mile End Road, LONDON, E1 4NS, U.K.}

\end{center}

\vspace{1in}

\baselineskip=24pt
\begin{abstract}
\noindent
It is shown that spatially flat, isotropic
cosmologies derived from the Brans--Dicke gravity  action exhibit
a scale factor duality invariance. This classical duality is then associated
with a hidden $N=2$ supersymmetry at the quantum level and the
supersymmetric quantum constraints are solved exactly.
These symmetries also apply to a dimensionally reduced version
of vacuum Einstein gravity.

\end{abstract}

\vspace{.7in}

PACS NUMBERS: 04.50.$+$h, 11.30.Pb, 98.80.Hw

\end{titlepage}


The possibility that scalar--tensor gravity
theories may be relevant in the arena of the very early Universe
has been widely investigated in recent years
\cite{scalartensor,BM1990,barrow,string,GV1993}. The defining feature of these
theories is the non--minimal coupling of a scalar field to
the space--time curvature. These couplings arise naturally in the low energy
limit of various unified  field theories such as superstring theory
\cite{string,GSW1988}. Theories containing higher--order terms in the
Ricci scalar may be also expressed in a scalar--tensor form by means of
a suitable conformal transformation \cite{W1993}. Furthermore,
the dimensional reduction of higher--dimensional gravity also
results in an effective scalar--tensor theory \cite{higher}.

The study of scalar--tensor theories is therefore well motivated. In this
paper we consider the $D$-dimensional, vacuum theory
\begin{equation}
\label{action}
S=\int d^D x\sqrt{-g} e^{-\Phi} \left[ R -\omega \left( \nabla \Phi
\right)^2 -2\Lambda \right]  ,
\end{equation}
where $\Phi$ represents the dilaton field,
$g$ is the determinant of the space-time metric ${\cal{G}}$, $\omega$
is a space--time constant and
$\Lambda$ is the effective cosmological constant.
This theory  represents one of the simplest extensions to Einstein gravity
and is formally equivalent to the standard
Brans-Dicke theory when $\Lambda =0$ \cite{BD1961}. It
represents the genus--zero effective action  for
the bosonic string when the  antisymmetric
tensor field $B_{\mu\nu}$ vanishes,
$\omega =-1$ and $\Lambda = (D-26)/3 \alpha'$, where $\alpha'$ is the inverse
string tension \cite{GSW1988}. The effective
action for the bosonic sector of the closed superstring also has this form,
but with $\Lambda$ proportional to $(D-10)$.

It is known that the reduced string action corresponding to cosmologies
that are spatially flat  and homogeneous exhibits a symmetry
known as {\em scale factor duality} \cite{dualref}.
The symmetry group is $Z^{D-1}_2$ and it is
generated by inverting the scale factor and shifting the value
of the dilaton. Scale factor duality is a special case of
a more general $O(d,d)$ symmetry of the theory and it
relates  expanding dimensions to contracting ones \cite{MV1991}.
It also  forms the basis for the  pre--big bang cosmological model developed
by Gasperini and Veneziano \cite{GV1993}.

In general, it is important to search for symmetries in a given
theory because they
can provide valuable insight into the dynamics of the Universe.
They  form the basis  for
selection rules that forbid  the existence of certain states and
processes. They also allow one to generate new, inequivalent
solutions to the field equations from a given solution. Moreover,
the conserved
quantity associated with a given symmetry represents an  integrability
condition on the field equations. In some cases, this may
allow more general solutions to be found.

In this paper we  find the analogue of
scale factor duality in theory (\ref{action}) when $\omega \ne - 1$.
We show that a duality invariance exists
for spatially flat, isotropic cosmologies of  arbitrary
dimension $D \ge 2$ when $\omega \ne -D/(D-1)$.
We then consider the quantization of this family of Universes by
mapping the system onto the constrained oscillator--ghost--oscillator
model. This procedure uncovers a hidden supersymmetry in the theory
that exists at the quantum level. The origin of this supersymmetry
may be traced to the scale factor duality invariance
of the classical action.
The super--quantum constraints,  corresponding to the Dirac-type
square root of the Wheeler-DeWitt
equation, are then solved exactly in closed form.

We express the space--time metric as
${\cal{G}}= {\rm diag} [ -N(t) ,G(t)]$,
where $t$ represents cosmic time, $N(t)$ is the lapse function
and the $(D-1)\times
(D-1)$ matrix $G (t)$ is the metric on the spatial hypersurfaces. We further
assume that the dilaton field is constant on these surfaces
and that $\omega > -(D-1)/(D-2)$. It then
follows that action (\ref{action}) reduces to
\begin{eqnarray}
\label{actionG}
S=\int dt e^{\ln \sqrt{{\rm det} G} -\Phi} \left[
\frac{1}{N} \left( 2 \partial^2_t \ln \sqrt{{\rm det} G} +\left(
\partial_t \ln \sqrt{{\rm det}G} \right)^2 \right. \right. \nonumber \\
\left. \left. -\frac{1}{4} {\rm Tr}
\left( \partial_t G \partial_t G^{-1} \right) + \omega (\partial_t \Phi)^2
\right) -2 N \Lambda \right] ,
\end{eqnarray}
where $\partial_t \equiv  \partial /\partial t$. The comoving
volume of the Universe has been normalized to unity in this expression
without loss of generality.
For the class of spatially flat, isotropic and homogeneous Universes,
the action simplifies further to
\begin{equation}
\label{isotropic}
S =  \int dt e^{(D-1) \alpha -\Phi} \left[ \frac{1}{N}
\left( -(D-1)(D-2) \dot{\alpha}^2 +2(D-1) \dot{\alpha}\dot{\Phi}
+\omega \dot{\Phi}^2 \right) -2N\Lambda \right]  ,
\end{equation}
where $e^{\alpha (t)}$ represents the scale
factor of the Universe, a dot denotes differentiation
with respect to $t$ and a boundary term has been neglected.
The field equations derived from this action take the form
\begin{eqnarray}
\label{field}
2\Lambda =2(D-2) \ddot{\alpha}  -2\ddot{\Phi} +(D-1)(D-2) \dot{\alpha}^2
+(2+\omega ) \dot{\Phi}^2  -2(D-2)\dot{\alpha}\dot{\Phi} \nonumber \\
2\Lambda = 2(D-1) \ddot{\alpha} +2\omega \ddot{\Phi} +D(D-1) \dot{\alpha}^2
-\omega \dot{\Phi}^2  +2(D-1) \omega \dot{\alpha} \dot{\Phi} \nonumber \\
2\Lambda = (D-1)(D-2) \dot{\alpha}^2 - \omega \dot{\Phi}^2
-2(D-1) \dot{\alpha}\dot{\Phi}
\end{eqnarray}
in the gauge $N=1$, where the third expression represents the
Hamiltonian constraint.

If we equate the first two expressions in Eq. (\ref{field})
and integrate, we find that
\begin{equation}
\label{conserved}
F \equiv e^{(D-1)\alpha -\Phi} \left[ \dot{\alpha} +(1+\omega ) \dot{\Phi}
\right]
\end{equation}
is a constant of motion, i.e., $\dot{F}=0$.
The conservation of this quantity is associated
with a non--trivial symmetry of the action
(\ref{isotropic}). It is known that this
action is invariant under time reversal $t=-\tilde{t}$.
However, it is also symmetric  under the simultaneous transformation
\begin{eqnarray}
\label{isodual}
\alpha =\left[ \frac{(D-2)+(D-1)\omega}{D+(D-1)\omega} \right]
\tilde{\alpha}
 - \left[ \frac{2(1+\omega)}{D+(D-1)\omega} \right] \tilde{\Phi} \nonumber \\
\Phi =- \left[ \frac{2(D-1)}{D+(D-1)\omega} \right] \tilde{\alpha} -\left[
\frac{(D-2)+(D-1) \omega}{D+(D-1)\omega} \right] \tilde{\Phi}
\end{eqnarray}
for all $D\ge 2$ and all $ \omega \ne  -D/(D-1)$.
It is straightforward to verify that the form of
$F$ remains invariant under these simultaneous interchanges.
The two--dimensional version of this symmetry  has been  discussed recently by
Cadoni and Cavagli\`a \cite{CC1994} and  Eq. (\ref{isodual}) reduces
to the scale factor duality invariance
of the string effective action when $\omega
=-1$. In this case, $\alpha = -\tilde{\alpha}$
and $\Phi = -2(D-1) \tilde{\alpha} + \tilde{\Phi}$ \cite{dualref}.

When $\omega \ne -D/(D-1)$,
Eqs. (\ref{field}) admit the exact solution $\Phi =vt$ and $\alpha =Ht$,
where $H=-(1+\omega )v$ and
\begin{equation}
\label{selfdual}
v^2 = \frac{2\Lambda}{[D-1 +(D-2)\omega][D+(D-1)
\omega]}  .
\end{equation}
This solution applies in the regime $\omega >-D/(D-1)$ when
$\Lambda >0$ and $\omega < -D/(D-1)$ when $\Lambda <0$.
It is an attractor to the general solution of Eqs. (\ref{field})
in the limit $t \rightarrow +\infty$
\cite{BM1990}. 	Application of Eq. (\ref{isodual}) maps
this attractor onto itself, so the solution is `self--dual'.
It is interesting that the symmetry
(\ref{isodual}) does not hold for $\omega =-D/(D-1)$ and
this feature appears to be
related to the form of the attractor. Consequently, we shall
not consider this  particular value further.

Before quantizing this model it proves convenient to express Eq.
(\ref{isotropic}) in a manifestly canonical form by means of an appropriate
change of variables. For $D\ge 3$,  we define
\begin{eqnarray}
\label{xy}
x\equiv  \exp \left[ \left( \frac{D-1}{2} + \frac{\gamma}{2} \right)
\left( \alpha + \frac{1}{D-2} \left( \frac{1}{\gamma} -1 \right)
\Phi \right) \right] \nonumber \\
y\equiv  \exp \left[ \left( \frac{D-1}{2} - \frac{\gamma}{2} \right)
\left( \alpha - \frac{1}{D-2} \left( \frac{1}{\gamma} +1 \right)
\Phi \right) \right]  ,
\end{eqnarray}
where
\begin{equation}
\label{gamma}
\gamma \equiv \left[ \frac{D-1}{D-1 + (D-2)\omega} \right]^{1/2}  .
\end{equation}
On the other hand, we define
\begin{eqnarray}
\label{2d}
x \equiv e^{\alpha + \omega \Phi /2} \nonumber \\
y \equiv e^{-(\omega +2)\Phi /2}
\end{eqnarray}
if $D=2$. Action (\ref{isotropic}) is therefore given by
\begin{equation}
\label{canonical}
S=\int dt \left[ -\frac{4}{N} \left(
\frac{D-1 +(D-2)\omega}{D+(D-1)\omega} \right) \dot{x}\dot{y} -2N\Lambda
xy \right]
\end{equation}
for all $D\ge 2$.

A second coordinate pair
\begin{eqnarray}
\label{wz}
w \equiv  \epsilon^{1/2}
\left[ \frac{D-1 +(D-2)\omega}{D+(D-1)\omega} \right]^{1/2} (x-y)
\nonumber \\
z \equiv \epsilon^{1/2}
\left[ \frac{D-1 +(D-2)\omega}{D+(D-1)\omega} \right]^{1/2} (x+y)
\end{eqnarray}
may now be introduced, where $\epsilon =+1$ if $\omega > -D/(D-1)$ and
$\epsilon =-1$ if $\omega < -D/(D-1)$.
It follows that action (\ref{canonical}) transforms
to
\begin{equation}
\label{quadratic}
S= \frac{1}{\epsilon}
\int  dt \left[ \frac{1}{N} \left( \dot{w}^2 -\dot{z}^2 \right)
-\frac{\lambda}{4}  \left( w^2 -z^2 \right) N \right] ,
\end{equation}
where
\begin{equation}
\lambda \equiv - 2\Lambda  \left[
\frac{D+(D-1)\omega}{D-1 +(D-2)\omega} \right]  .
\end{equation}
This is the action for the constrained
oscillator--ghost--oscillator pair  when $\lambda >0$ \cite{note}.  The pair
oscillate with identical frequency, but have equal and opposite energy.
The general solution to the classical field equations of this model
may be represented by a family of ellipses
in the $(w,z)$ plane \cite{OT1994}.
The major axis of the trajectories is given by $w=\pm z$ and the
eccentricities are determined by the  integration constants. Physical
solutions for the scalar--tensor action (\ref{isotropic})  are
restricted to the $|w| \le z$ sector of the plane.

Rewriting the system in terms of these new variables is useful
because the scale factor duality invariance of action (\ref{isotropic})
becomes more apparent. Substitution of Eq.  (\ref{isodual}) into
Eqs. (\ref{xy}) and (\ref{2d}) implies that the
duality transformation is formally equivalent to the simultaneous interchange
of the canonical variables  $x \leftrightarrow y$. Action (\ref{canonical})
is clearly symmetric under this interchange. Moreover, it then
follows directly from the definition (\ref{wz}) that
the transformation (\ref{isodual})
may also be generated by  $w \rightarrow -w$ and $z\rightarrow z$.

The classical Hamiltonian for this system is given by
\begin{equation}
\label{classicalham}
2H_0= G^{\mu\nu} p_{\mu}p_{\nu} +W(q^{\mu}) =0  ,
\end{equation}
where $\mu ,\nu =0,1$, and $G^{\mu\nu} ={\rm diag} [-1/2,1/2]$ is
proportional to the inverse of the metric over the configuration
space. This space is spanned by the `time--like' coordinate $q^0=w$ and
`space--like' coordinate $q^1=z$. The momenta conjugate
to these variables are  $p_0 = \partial S/\partial q^0 = 2\dot{w}/N$
and $p_1 =\partial S /\partial
q^1 =-2\dot{z}/N$, respectively, and the potential is given by
\begin{equation}
\label{super}
W= - \lambda (w^2 -z^2) /2  .
\end{equation}

In the standard approach to quantum cosmology, one views
$H_0$ as an operator that annihilates the
state  vector $\Psi$ of the Universe, i.e., $H_0 \Psi =0$.
This is the  Wheeler-DeWitt equation \cite{W1967}.
To quantize the model  one imposes
the algebra $\left[ q^{\mu} ,p^{\nu} \right]_- =i  \delta^{\mu\nu}$
and this is realized  by identifying $p^{\mu} =-i \partial_{\mu}$
$(\hbar =1 )$.
If we neglect any ambiguities that may arise  due to factor
ordering, the Wheeler-DeWitt equation takes the form
\begin{equation}
\label{WDW}
\left[ \frac{\partial^2}{\partial w^2} -\frac{\partial^2}{\partial z^2}
-\lambda (w^2 -z^2) \right] \Psi =0
\end{equation}
and admits the family of solutions
\begin{equation}
\label{descrete}
\Psi_n = H_n ( \lambda^{1/4} w  ) H_n ( \lambda^{1/4} z )
e^{-\sqrt{\lambda}  (w^2+z^2)/2}  ,
\end{equation}
where $H_n$ is the Hermite polynomial of order $n$.
When $\lambda >0$, these solutions
form a descrete basis for any bounded wavefunction $\Psi = \sum
c_n \Psi_n$, where $c_n$ are complex coefficients \cite{OT1994,P1991,other}.
The ground state corresponds to $n=0$ and excited states to
$n > 0$. The ground state is symmetric
under the action of Eq. (\ref{isodual}).
These states do not oscillate and therefore represent
Euclidean geometries \cite{H1984}. Oscillating wavefunctions
representing Lorenztian geometries may be found, however,  when the $c_n$
take appropriate values \cite{OT1994}.

Further insight may be gained by introducing new variables
$u = \sqrt{\lambda} ({w} +{z})^2/4$ and $v = \sqrt{\lambda} ({w} -{z})^2/4$.
This transforms the Wheeler-DeWitt equation (\ref{WDW}) into the
unit--mass Klein--Gordon  equation. One family of solutions
to this equation is given by $\Psi_b = e^{-bu-v/b}$, where $b$ is an
arbitrary, complex constant and $|\Psi_b|$ is bounded for Re$b >0$.
This family represents  the basis
for a continuous spectrum  of states of the form
$\Psi = \int_C db M(b) \Psi_b$, where $M(b)$ is an arbitrary function of
$b$ and $C$ is some contour in the Re$b>0$ sector of the complex $b$
plane \cite{P1991}.
Since the $\Psi_{b=1}$ solution coincides with the ground
state $\Psi_{n=0}$
of the descrete spectrum (\ref{descrete}), it may be viewed as the
ground state of this continuous spectrum. The excited states
therefore correspond to $b\ne 1$.
Moreover, the transformation (\ref{isodual}) is equivalent
to the simultaneous interchange $u \leftrightarrow v$. Thus, the ground state
of the continuous  spectrum respects the symmetry of the action
(\ref{isotropic}), but the excited states do not. In this sense, therefore,
the ground state represents the maximally symmetric, self--dual  solution.

We shall now illustrate how the invariance of action (\ref{isotropic})
under Eq. (\ref{isodual}) implies that the classical
Hamiltonian $H_0$  is the bosonic component of a supersymmetric
Hamiltonian when $\lambda >0$.
In general, there exists a hidden supersymmetry in the theory
if there is a solution to the
Euclidean Hamilton--Jacobi equation that respects
the same symmetries as $H_0$ \cite{super}.
The Euclidean Hamilton--Jacobi  equation has the form
\begin{equation}
\label{HJ}
G^{\mu\nu} \frac{\partial I}{\partial q^{\mu}} \frac{\partial
I}{\partial q^{\nu}} =W  ,
\end{equation}
where $I=I(q^{\mu})$ and,  for the model we are
considering here, admits the exact solution
\begin{equation}
\label{I}
I=\sqrt{\lambda} (w^2+z^2) /2  .
\end{equation}
This solution is manifestly invariant under
the duality transformation $w\rightarrow -w$ and $z\rightarrow z$.

The system with Hamiltonian (\ref{classicalham}) may now be quantized
by considering the quantum Hamiltonian $H$ defined by
\begin{equation}
\label{superham}
2H = [ Q,\bar{Q} ]_+ ,\qquad Q^2 = \bar{Q}^2 =0  ,
\end{equation}
where $Q$ is a non--Hermitian supersymmetry charge with adjoint $\bar{Q}$.
It follows that $[H,Q]_- =[H,\bar{Q}]_- =0$ and
Eq. (\ref{superham}) represents the algebra for an $N=2$
supersymmetry \cite{super,G1991}.
The supercharges are linear operators of the form
\begin{equation}
\label{Q}
Q\equiv \varphi^{\mu} \left( p_{\mu} +i \frac{\partial I}{\partial q^{\mu}}
\right) ,\qquad \bar{Q} \equiv \bar{\varphi}^{\mu} \left(
p_{\mu} -i \frac{\partial I}{\partial q^{\mu}} \right)  ,
\end{equation}
where the fermionic degrees of freedom $\varphi^{\mu}, \bar{\varphi}^{\nu}$
satisfy the spinor algebra
\begin{equation}
\label{spinoralgebra}
\left[ \varphi^{\mu},\varphi^{\nu} \right]_+ =\left[
\bar{\varphi}^{\mu},\bar{\varphi}^{\nu} \right]_+ =0 , \qquad
\left[ \varphi^{\mu},\bar{\varphi}^{\nu} \right]_+ =G^{\mu\nu}
\end{equation}
and the bosonic degrees of freedom satisfy the usual anticommutation
relation. It follows, therefore,  that the quantum Hamiltonian has the form
\begin{equation}
\label{quantumham}
H=H_0  +\frac{\hbar}{2} \frac{\partial^2 I}{\partial q^{\mu}
\partial q^{\nu}} \left[ \bar{\varphi}^{\mu} ,\varphi^{\nu} \right]_-
\end{equation}
and reduces to the bosonic Hamiltonian
(\ref{classicalham}) in the classical limit. However, it acquires an
additional spin term at the quantum level whose form  suggests that imaginary
solutions to Eq. (\ref{HJ}) will be inappropriate. In view of this
we restrict the analysis to the region of parameter space where $\lambda >0$.
This corresponds to $\omega < -D / (D-1)$ when $\Lambda >0$ and
$\omega >-D/(D-1)$ if $\Lambda <0$.
We note that $\lambda >0$ for the effective
bosonic (super--) string action if $D<26$ $(D<10)$. The
existence of a hidden supersymmetry in the two--dimensional bosonic string
cosmology has recently been discussed in Ref. \cite{L1995}.

A suitable representation for the fermionic sector
of the supersymmetric  algebra is given
in terms of the Grassmann variables $\theta^{\mu}$ and their derivatives,
i.e., $\bar{\varphi}^{\mu} = \theta^{\mu}$ and $\varphi^{\mu} = G^{\mu\rho}
\partial  /\partial \theta^{\rho}$. The general form of the
supersymmetric wavefunction may then be expanded
in terms of these variables such that $\Psi =
A_+ +B_0\theta^0 +B_1 \theta^1 +A_- \theta^0 \theta^1$, where the
coefficients $A_{\pm}$, $B_0$ and $B_1$
are functions of $(w,z)$ only \cite{BG1994}.
This wavefunction is annihilated by the supercharges
$Q\Psi =\bar{Q} \Psi =0$
and it is these constraints that represent the Dirac--type square root
of the Wheeler-DeWitt equation. They are given by a set of coupled, linear
differential equations:
\begin{eqnarray}
\left[ \frac{\partial}{\partial w} +\frac{\partial I}{\partial w}
\right] A_+=0
\nonumber \\
\left[ \frac{\partial}{\partial z} +\frac{\partial I}{\partial z}
\right] A_+=0
\nonumber \\
\left[ \frac{\partial}{\partial w} +\frac{\partial I}{\partial w}
\right]
B_1 -\left[ \frac{\partial}{\partial z} +
\frac{\partial I}{\partial z} \right]
B_0=0 \nonumber \\
\left[ \frac{\partial}{\partial w} -\frac{\partial I}{\partial w}
\right]
B_0 -\left[ \frac{\partial}{\partial z} -\frac{\partial I}{\partial z}
\right] B_1=0 \nonumber \\
\left[ \frac{\partial}{\partial w} -\frac{\partial I}{\partial w}
\right]
A_- =0 \nonumber \\
\left[ \frac{\partial}{\partial z} -\frac{\partial I}{\partial z}
\right] A_- =0 ,
\end{eqnarray}
where $I$ is given by Eq. (\ref{I}).

These equations may  be solved exactly and the supersymmetric
wavefunction has the form
\begin{eqnarray}
\label{supersol}
\Psi = e^{-I} + 2n\lambda^{1/4} \left[ H_{n-1} \left( \lambda^{1/4} w
\right) H_n \left( \lambda^{1/4} z \right)  \theta^0  \right. \nonumber \\
\left. +
H_n \left( \lambda^{1/4} w \right) H_{n-1} \left( \lambda^{1/4} z \right)
\theta^1 \right] e^{-I}
+e^I \theta^0 \theta^1  ,
\end{eqnarray}
where $n\ge 0$.
The solutions $A_{\pm}$ represent the empty and filled fermion
sectors of the wavefunction, respectively. As can be seen directly
from Eq. (\ref{supersol}), they are manifestly
self--dual. Moreover, the empty fermion sector of this wavefunction
is identical to the ground states $\Psi_{n=0}$ and $\Psi_{b=1}$
of the purely bosonic spectra
discussed above. Hence, this supersymmetric approach to
quantum cosmology naturally selects the bosonic state of lowest energy.

In summary, we have found that a class of scalar--tensor
cosmologies -- including the vacuum Brans--Dicke model -- exhibit
a scale factor duality invariance. By mapping the
model onto the zero--energy oscillator--ghost--oscillator pair, we
identified a continuous spectrum of quantum states whose ground
state is the self--dual wavefunction. We
further showed that this duality symmetry of the classical action
is associated with a hidden supersymmetry
that exists at the quantum level in a wide region of parameter space.
Included in this regime is the bosonic string effective action.
This is important because supersymmetric quantum cosmology may be able to
resolve the problems encounted in the standard approach when one attempts
to construct  a non--negative norm from the wavefunction \cite{BG1994}.
Furthermore, Dereli, \"Onder and
Tucker \cite{DOT1994} have developed an alternative  spinor model of quantum
cosmology that is also based on the constrained oscillator--ghost--oscillator.
The direct correspondence between Eqs. (\ref{isotropic}) and (\ref{quadratic})
suggests  that their results will also apply for the model discussed here
and it would be of interest to compare and contrast
these two approaches in more detail.

It is also of interest to investigate  whether the scale factor duality
invariance of action (\ref{action})
can be extended to more general space--times.
The simplest extensions are to include spatial curvature and anisotropy.
However,
spatial curvature introduces  a term proportional to $e^{-2\alpha}$ into the
Lagrangian of action (\ref{isotropic}) and this
explicitly breaks the symmetry of the  model. Thus, within
the context of spatially isotropic Universes, the
duality symmetry and associated hidden supersymmetry of action (\ref{action})
are only respected  by  the spatially flat model.

If the Universe is anisotropic, however,
with a line element given by $ds^2=-dt^2 +e^{2\alpha_i (t)}dx_i^2$,
Eq. (\ref{action}) takes the form
\begin{equation}
\label{anisotropic}
S=\int dt e^{\left( \sum_i \alpha_i -\Phi \right)}  \left[
 2 \dot{\Phi} \sum_{i} \dot{\alpha}_i -
\sum_{i,j}\dot{\alpha}_i \dot{\alpha}_j
+ \sum_i \dot{\alpha}^2_i +\omega \dot{\Phi}^2  -2\Lambda
\right]  ,
\end{equation}
where the summations are over $i,j=1, \ldots , (D-1)$.
We may now consider the generalization of transformation (\ref{isodual})
given by
\begin{eqnarray}
\label{anisduality}
\alpha_i =c_1 \tilde{\alpha}_i +c_2 \tilde{\Phi} \nonumber \\
\Phi = c_3 \sum_{i=1}^{D-1} \tilde{\alpha}_i +c_4 \tilde{\Phi}  ,
\end{eqnarray}
where the $c_l$'s are arbitrary constants. Substitution of Eq.
(\ref{anisduality}) into Eq. (\ref{anisotropic}) leads to six
constraints that must be simultaneously satisfied if the action is to remain
invariant. However, the only non-trivial solution to these constraints
is given by $c_1=-1$, $c_2=0$, $c_3=-2$, $c_4=1$ and $\omega=-1$.
Thus,  the scale factor duality invariance
(\ref{anisduality}) is unique to the string effective
action if an anisotropic Universe is considered.
If $\omega  \ne -1$, the symmetry only
holds for spatially flat, isotropic models.

Although these conclusions apply to cosmologies derived
from the scalar--tensor action (\ref{action}), they
are also relevant to the higher--dimensional, vacuum Einstein action
$S=\int d^T x \sqrt{-g} \left[ R - 2 \Lambda \right]$, where $T >D$.
In general, scale factor duality maps a constant gravitational coupling onto
a time--dependent one. Consequently, it
is not respected by Einstein gravity when the Universe is spatially isotropic.
However, it can be if  a suitable dimensional reduction of the theory
is considered. If the higher--dimensional space--time  $M$
is viewed as the  product $M = R\times
J \times K$, where $J$ is the  $(D-1)$--dimensional
`external' space  and $K$ is a Ricci--flat, isotropic  $d$--dimensional space,
the dimensionally reduced action has the form of Eq. (\ref{action}),
where $\omega =-1+1/d$ and $\Phi$ is related to the radius of $K$.
Since $\omega >-1$ in this model for  all physical values of $d$, it follows
from the above discussion that the dimensionally reduced
action will not be invariant under
Eq. (\ref{anisduality}) if the external space is
anisotropic. On the other hand, the action is invariant under
the classical scale factor duality (\ref{isodual}) if
$J$ is  isotropic and spatially flat. In this case, one may then
verify that there  exists a hidden
$N=2$ supersymmetry for all values of $D$ and $d$ if $\Lambda <0$,
but not if $\Lambda >0$.

In conclusion, therefore,
we have identified  a dimensionally reduced
version of the higher--dimensional, vacuum Einstein action that is invariant
under a  scale factor duality  transformation. This symmetry
is respected if the higher--dimensional Universe is anisotropic
with a spatial geometry given by the
product of two spaces that are both flat and isotropic.
The symmetry discussed in this work is a
natural generalization of the scale
factor duality exhibited by the genus--zero string effective action.
If the cosmological constant is negative, it may be further
associated with a hidden $N=2$ supersymmetry.

\vspace{1cm}

The author is supported by the Particle Physics and Astronomy Research
Council (PPARC), UK. We thank J. D. Barrow and J. Garc\'ia--Bellido
for helpful discussions.

\end{document}